\PassOptionsToPackage{unicode}{hyperref}
\PassOptionsToPackage{hyphens}{url}
\PassOptionsToPackage{dvipsnames,svgnames,x11names}{xcolor}
\documentclass[
  12pt]{article}
\usepackage{comment}
\usepackage{amsmath,amssymb}
\usepackage{iftex}
\ifPDFTeX
  \usepackage[T1]{fontenc}
  \usepackage[utf8]{inputenc}
  \usepackage{textcomp} 
\else 
  \usepackage{unicode-math}
  \defaultfontfeatures{Scale=MatchLowercase}
  \defaultfontfeatures[\rmfamily]{Ligatures=TeX,Scale=1}
\fi
\usepackage{lmodern}
\ifPDFTeX\else  
\fi
\IfFileExists{upquote.sty}{\usepackage{upquote}}{}
\IfFileExists{microtype.sty}{
  \usepackage[]{microtype}
  \UseMicrotypeSet[protrusion]{basicmath} 
}{}
\makeatletter
\@ifundefined{KOMAClassName}{
  \IfFileExists{parskip.sty}{%
    \usepackage{parskip}
  }{
    \setlength{\parindent}{0pt}
    \setlength{\parskip}{6pt plus 2pt minus 1pt}}
}{
  \KOMAoptions{parskip=half}}
\makeatother
\usepackage{xcolor}
\makeatletter
\ifx\paragraph\undefined\else
  \let\oldparagraph\paragraph
  \renewcommand{\paragraph}{
    \@ifstar
      \xxxParagraphStar
      \xxxParagraphNoStar
  }
  \newcommand{\xxxParagraphStar}[1]{\oldparagraph*{#1}\mbox{}}
  \newcommand{\xxxParagraphNoStar}[1]{\oldparagraph{#1}\mbox{}}
\fi
\ifx\subparagraph\undefined\else
  \let\oldsubparagraph\subparagraph
  \renewcommand{\subparagraph}{
    \@ifstar
      \xxxSubParagraphStar
      \xxxSubParagraphNoStar
  }
  \newcommand{\xxxSubParagraphStar}[1]{\oldsubparagraph*{#1}\mbox{}}
  \newcommand{\xxxSubParagraphNoStar}[1]{\oldsubparagraph{#1}\mbox{}}
\fi
\makeatother

\usepackage{longtable,booktabs,array}
\usepackage{calc} 
\usepackage{etoolbox}
\makeatletter
\patchcmd\longtable{\par}{\if@noskipsec\mbox{}\fi\par}{}{}
\makeatother
\IfFileExists{footnotehyper.sty}{\usepackage{footnotehyper}}{\usepackage{footnote}}
\makesavenoteenv{longtable}
\usepackage{graphicx}
\makeatletter
\def\maxwidth{\ifdim\Gin@nat@width>\linewidth\linewidth\else\Gin@nat@width\fi}
\def\maxheight{\ifdim\Gin@nat@height>\textheight\textheight\else\Gin@nat@height\fi}
\makeatother
\setkeys{Gin}{width=\maxwidth,height=\maxheight,keepaspectratio}
\makeatletter
\def\fps@figure{htbp}
\makeatother

\makeatletter
\@ifpackageloaded{caption}{}{\usepackage{caption}}
\AtBeginDocument{%
\ifdefined\contentsname
  \renewcommand*\contentsname{Table of contents}
\else
  \newcommand\contentsname{Table of contents}
\fi
\ifdefined\listfigurename
  \renewcommand*\listfigurename{List of Figures}
\else
  \newcommand\listfigurename{List of Figures}
\fi
\ifdefined\listtablename
  \renewcommand*\listtablename{List of Tables}
\else
  \newcommand\listtablename{List of Tables}
\fi
\ifdefined\figurename
  \renewcommand*\figurename{Figure}
\else
  \newcommand\figurename{Figure}
\fi
\ifdefined\tablename
  \renewcommand*\tablename{Table}
\else
  \newcommand\tablename{Table}
\fi
}
\@ifpackageloaded{float}{}{\usepackage{float}}
\floatstyle{ruled}
\@ifundefined{c@chapter}{\newfloat{codelisting}{h}{lop}}{\newfloat{codelisting}{h}{lop}[chapter]}
\floatname{codelisting}{Listing}

\makeatother
\makeatletter
\@ifpackageloaded{caption}{}{\usepackage{caption}}
\@ifpackageloaded{subcaption}{}{\usepackage{subcaption}}
\makeatother

\ifLuaTeX
  \usepackage{selnolig}  
\fi
\usepackage[]{natbib}
\usepackage{bookmark}

\IfFileExists{xurl.sty}{\usepackage{xurl}}{} 
\hypersetup{
  pdftitle={Title},
  pdfauthor={Author 1; Author 2},
  pdfkeywords={3 to 6 keywords, that do not appear in the title},
  colorlinks=true,
  linkcolor={blue},
  filecolor={Maroon},
  citecolor={Blue},
  urlcolor={Blue},
  pdfcreator={LaTeX via pandoc}}

\newcommand{\bA}{\mathbf{A}}
\newcommand{\bbR}{\mathbb{R}}
\newcommand{\bbE}{\mathbb{E}}

\newcommand{\bI}{\mathbf{I}}

\newcommand{\bB}{\mathbf{B}}
\newcommand{\bX}{\mathbf{X}}

\newcommand{\bW}{\mathbf{W}}

\newcommand{\bF}{\mathbf{F}}
\newcommand{\vf}{\mathbf{f}}

\newcommand{\bx}{\mathbf{x}}

\newcommand{\bz}{\mathbf{z}}

\newcommand{\bS}{\mathbf{S}}
\newcommand{\bT}{\mathbf{T}}
\newcommand{\bu}{\mathbf{u}}
\newcommand{\bU}{\mathbf{U}}

\newcommand{\bw}{\mathbf{w}}

\newcommand{\balpha}{\boldsymbol{\alpha}}
\newcommand{\bbeta}{\boldsymbol{\beta}}

\newcommand{\btheta}{\boldsymbol{\theta}}

\newcommand{\bphi}{\boldsymbol{\phi}}
\newcommand{\bPhi}{\boldsymbol{\Phi}}

\newcommand{\bSigma}{\boldsymbol{\Sigma}}

\newcommand{\bgamma}{\boldsymbol{\gamma}}
\newcommand{\bGamma}{\boldsymbol{\Gamma}}
\newcommand{\bLambda}{\boldsymbol{\Lambda}}
\newcommand{\bxi}{\boldsymbol{\xi}}

\newcommand{\bzero}{{\bf 0}}

\def\argmin{\mathop{\rm arg\ min}\limits}
\def\argmax{\mathop{\rm arg\ max}\limits}

\newtheorem{theorem}{Theorem}

\newtheorem{case}{Case}

\newtheorem{lemma}{Lemma}

\newtheorem{coro}{Corollary}
\newtheorem{asmp}{Assumption}

\newcommand{\anon}{1}

\begin{document}

\def\spacingset#1{\renewcommand{\baselinestretch}%
{#1}\small\normalsize} \spacingset{1}


\if1\anon
{
  \title{\bf Conditional inference for high-dimensional multi-omics survival data}
  \author{Heyuan Zhang \\
    \small{SKLMS, Academy of Mathematics and Systems Science, Chinese Academy of Sciences, China}\\
    \small{School of Mathematical Sciences, University of Chinese Academy of Sciences, China}\\
  Meiling Hao \\
    \small{School of Statistics, University of International Business and Economics, China}\\
    Lianqiang Qu\\
    \small{School of Mathematics and Statistics, Central China Normal University,  China}\\
    Liuquan Sun\\
    \small{SKLMS, Academy of Mathematics and Systems Science, Chinese Academy of Sciences, China}\\
\small{School of Mathematical Sciences, University of Chinese Academy of Sciences, China}}
   \date{}
  \maketitle
} 
\fi

\if0\anon
{
  \bigskip
  \bigskip
  \bigskip
  \begin{center}
    {\LARGE\bf Conditional inference for high-dimensional multi-omics survival data}
\end{center}
  \medskip
} \fi

\bigskip
\begin{abstract}
Multi-omics data present significant challenges for statistical inference due to the complex interdependencies among biological layers. 
In this paper, we introduce a novel Multi-Omics Factor-Adjusted Cox (MOFA-Cox) model for analyzing multi-omics survival data, 
effectively addressing the intricate correlations across various omics layers.
We provide a factor-adjusted decorrelated score test for the MOFA-Cox model in high-dimensional survival analysis. 
Our method accommodates situations where the dimension of the parameters being tested exceeds the sample size, 
while not imposing a sparsity assumption on them.
We establish the limiting null distribution of the proposed test and analyze its power under local alternatives. 
Numerical studies and an application to the TCGA breast cancer dataset demonstrate the effectiveness of our method.
\end{abstract}

\noindent%
\vfill

\newpage
\spacingset{1.8} 

\renewcommand{\theequation}{\thesection.\arabic{equation}}	
	
	\setcounter{equation}{0}
	
\section{Introduction}\label{sec:intro}

Multi-omics data analysis facilitates a comprehensive and systematic understanding of biological mechanisms and complex systems by integrating information across multiple omics layers, including genomics, transcriptomics, proteomics, and metabolomics. Such integration can reveal biological insights that remain inaccessible through single-omics approaches, particularly in elucidating disease mechanisms \citep{clark2021integrative, chen2023applications} and identifying potential biomarkers \citep{olivier2019need, garg2024disease}.

As a motivating example, we consider the TCGA breast cancer (BRCA) dataset, which categorizes gene expression levels into two omics layers: Protein-Coding and Long Non-Coding RNA (lncRNA). 
Breast cancer is a complex disease and remains one of the most prevalent cancers  affecting women worldwide.
Increasing evidence indicates that breast cancer is closely associated with various omics data. 
For instance, a purine derivative, QS11, has been shown to inhibit the migration of ARFGAP overexpressing breast cancer cells \citep{zhang2007small}. 
In the current study, our aim is to investigate whether lncRNA expression significantly impacts breast cancer while accounting for protein-coding gene effects.
This necessitates developing conditional inference methods, explicitly capturing dependencies between omics layers \citep{chen2023testing,yang2024score}.

\begin{figure}[thp]
    \centering
    \begin{subfigure}[b]{0.3\textwidth}
        \centering
        \includegraphics[width=\textwidth]{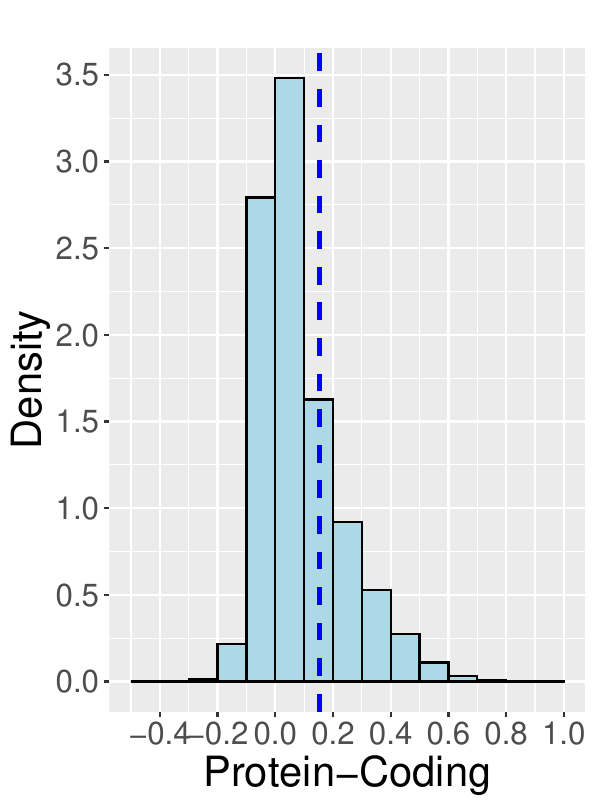}
    \end{subfigure}  
    \hspace{0.01\textwidth}
    \begin{subfigure}[b]{0.3\textwidth}
        \centering
        \includegraphics[width=\textwidth]{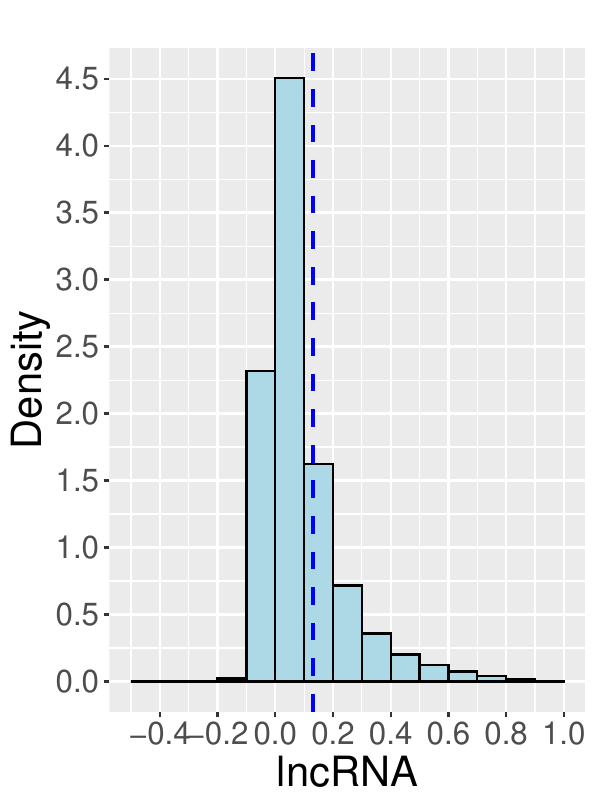}
    \end{subfigure}
    \hspace{0.01\textwidth}
    \begin{subfigure}[b]{0.3\textwidth}
        \centering
        \includegraphics[width=\textwidth]{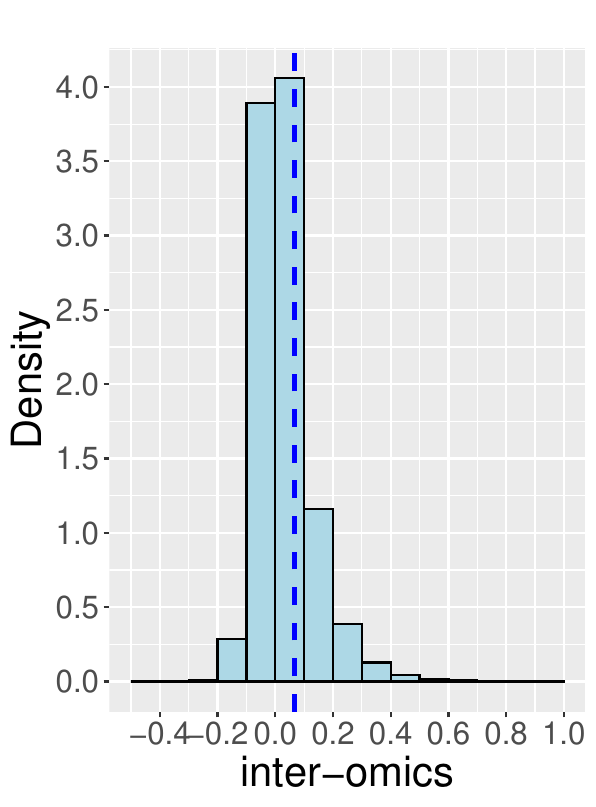}
    \end{subfigure}

    \caption{Histogram of  intra- and inter-omics sample correlation coefficient with the blue dashed line which shows the 25\% upper-quantile.}
    \label{fig3}
\end{figure}

The BRCA dataset includes expression levels for 19,962 protein-coding genes and 16,901 lncRNA genes, far exceeding the number of samples ($n = 1111$). This high-dimensionality places the analysis within the ``large $p$, small $n$'' framework. To illustrate the dependency structure across omics, Figure~\ref{fig3} displays the  intra- and inter-omics correlations.
 The left and middle panels show that both protein-coding and lncRNA genes exhibit strong intra-omics correlations. The right panel highlights substantial inter-omics correlations as well. 
Such strong correlations stem from latent biological processes and shared regulatory mechanisms across omics. They pose significant challenges for statistical modeling, especially in estimating the effects of individual features on clinical outcomes, which are often subject to right censoring.

The Cox model \citep{cox1972} is widely used in survival analysis. In the context of high-dimensional survival data analysis, 
various penalized methods, such as LASSO \citep{lasso1997} and SCAD \citep{fan2002}, have been developed to select important covariates. 
\cite{lasso2013} and \cite{lasso2014} established oracle inequalities for LASSO under the sparsity assumptions. 
Furthermore, \cite{Bradic_2011} demonstrated strong oracle properties of nonconcave penalized
methods for the Cox model; however, it requires irrepresentable conditions and ignores the uncertainty of model selection. 
\cite{zhong2015} considered conditional inference for additive hazard models.
\cite{jrssb2017} developed a unified likelihood ratio inferential framework for  proportional hazards models.
\cite{zhang2022projection} proposed a projection-based cross-validation method.
However, these methods are  tailored to test low-dimensional components within high-dimensional models, 
which limits their applicability to multi-omics data. Given the ultra-high dimensionality and complex dependence structures within and across omics layers, 
direct application of the aforementioned methods often leads to inflated Type I error. 
Recently, \cite{li2021integrative} proposed an integrative linear regression model using latent factors for multi-omics data. 
\cite{liWang} considered the factor-adjusted methods in generalized linear regression models.
However, their method is not applicable to right-censored outcomes, which are common in survival analysis.

To overcome these issues, we propose a new Multi-Omics Factor-Adjusted Cox (MOFA-Cox) model  for the analysis of multi-omics survival data.
Specifically, we adopt a factor analysis framework in which covariates are modeled as functions of common latent factors and uncorrelated idiosyncratic components, enabling effective variance decomposition to address complex correlation structures.
We then develop a conditional inference procedure based on decorrelated score tests, 
which are applied separately to each omics layer and tailored for high-dimensional features.
We derive the asymptotic distribution of the proposed test statistic
and investigate the power under a local alternative hypothesis.
The incorporation of estimated factor effects preserves consistency under the null hypothesis and increases power under the alternative.
Furthermore,  we address a non-typical Dantzig problem to achieve consistent estimation of the projection matrix, effectively overcoming computational challenges.

In essence, our approach differs from that of \cite{li2021integrative} in two fundamental aspects. 
First, their inference procedure relies on the assumption that 
the objective function can be expressed as a sum of independent and identically distributed (i.i.d.) components, 
which greatly simplifies theoretical analysis. 
In contrast, our MOFA-Cox model is based on counting processes that do not exhibit such an i.i.d. structure, 
introducing substantial technical challenges in the derivation of asymptotic properties. 
Second, while their test statistic exploits the linear model structure by projecting latent factors of the target group onto a subspace spanned by the other groups, 
our method constructs a projection based on the score function. 
This leads to the fact that many cross-product terms cannot be canceled out, 
further complicating the theoretical analysis. 
Thus, within the counting process framework, it requires developing new techniques to establish the asymptotic distribution of our test statistic.
The details of the differences and challenges compared with existing methods are provided in Section~\ref{asymptotic}.

The following notations are used throughout this article. Let $\mathbb{I}()$ denote the indicator function.
For a vector $\balpha=(\alpha_1,\dots,\alpha_p)^\top\in \mathbb{R}^p$, define $\|\balpha\|_q=(\sum_{i=1}^p|\alpha_i|^q)^{1/q}$ for $q\ge1$, $\|\balpha\|_\infty=\max_{1\le i\le p}|\alpha_i|$, and $\|\balpha\|_0=\sum_{i=1}^p\mathbb{I}(\alpha_i\neq 0)$. For any integer $p$, we denote $[p]=\{1,\dots,p\}$. 
For an index set $\mathcal{S}\subset [p]$, let $\balpha_{\mathcal{S}}$ denote the subvector of $\balpha$ corresponding to the indices in $\mathcal{S}$. 
Define $\mathbf 1_p=(1,\dots,1)^\top\in \mathbb{R}^{p}$, $\mathbf 0_p=(0,\dots,0)^\top\in \mathbb{R}^{p}$, and $\bI_p$ as the identity matrix in $\mathbb{R}^{p\times p}$.
For a matrix $\bA=(a_{ij}, 1\le i, j\le p)$, define $\|\bA\|_1=\sum_{1\le i,j\le p}|a_{ij}|,~\|\bA\|_\infty=\max_{1\le i,j\le p}|a_{ij}|$ and $\|\bA\|_F=(\sum_{1\le i, j\le p}a_{ij}^2)^{1/2}$ 
as its element-wise sum-norm, element-wise max-norm and Frobenius norm, respectively. 
Let $\lambda_{\min}(\bA)$ and $\lambda_{\max}(\bA)$ represent the minimal and maximal eigenvalues of $\bA$, respectively.
Furthermore, define $\|\bA\|_2=\lambda^{1/2}_{\max}(\bA^\top\bA)$. 
For any two positive sequences $\{a_n\}_{n\ge 1}$ and $\{b_n\}_{n\ge 1}$, 
we write $a_n\lesssim b_n$ if $a_n=O(b_n)$, and $a_n\asymp b_n$ if $a_n\lesssim b_n$ and $b_n\lesssim a_n$. 
In addition, for any two random sequences, we write $a_n=O_{p}(b_n)$ if $\lim_{n\to\infty}\mathbb P(|a_n/b_n|\le C)=1$ for some positive constant $C$,  
and $a_n=o_{p}(b_n)$ if $\lim_{n\to\infty}\mathbb P(|a_n/b_n|> C)=0$ for any positive constant $C$. 
For a set $\mathcal{S}$, let $|\mathcal{S}|$ denote the cardinality of $\mathcal{S}$. Let $\nabla$ denote the derivative operator.

The rest of the article is organized as follows. In Section~\ref{sec:pre}, we propose the MOFA-Cox model.
In Section \ref{sec:FADS}, we develop a factor-adjusted decorrelated score test for conditional inference. In Section~\ref{asymptotic}, we establish the asymptotic distribution under both the null and local alternative hypotheses. Section~\ref{simulation} focuses on simulation studies, while Section~\ref{data} demonstrates the application of our method to a real data example. Conclusions and discussions are provided in Section~\ref{discussion}. The proofs of the asymptotic properties are included in the supplementary materials.

\section{MOFA-Cox model}\label{sec:pre}
Let \( T \) and \( C \) denote the survival and censoring times, respectively. Define \( Y =\min(T,C) \) as the observed time and \( \delta = \mathbb{I}(T \leq C) \) as the censoring indicator. 
Suppose that there are \( M \) groups of covariates. Let \( \bx_m \in \mathbb{R}^{p_m} \) be the \( m \)-th group of covariates and define \( \mathcal{X} = (\bx_1^\top, \dots, \bx_M^\top)^\top \in \mathbb{R}^p \), where \( p = \sum_{m=1}^{M} p_m \).  
 For $t\in[0,\tau]$, define the counting process $N(t)=\mathbb{I}(Y\le t,\delta=1)$ and the risk indicator process $Y(t)=\mathbb{I}(Y\ge t)$, where $\tau<\infty$ is the maximum follow-up time.
        
        Under the Cox model, the hazard function of the survival time $T$ takes the form:
	\begin{align}\label{cox}
		\lambda(t|  \mathcal{X})=\lambda_0(t)\exp\left(\sum_{m=1}^M\bx_m^\top\bbeta_m^*\right),
	\end{align}
        where $\lambda_0(t)$ is the baseline hazard function and \( \bbeta_m^* \) denotes the coefficient of the \( m \)-th covariate group \( \bx_m \).

        For the omics analysis, we aim to test whether covariates in the $m$-th group exhibit significant effects on the hazard of $T$ when conditioned on all other groups. 
        We formulate this issue as the following hypothesis:
	\begin{align}\label{2.3}
		H_0: \boldsymbol{\beta}_m^*=\mathbf{0} \quad \text { versus } \quad H_a: \boldsymbol{\beta}_m^* \neq \mathbf{0}.
	\end{align}
Directly testing the null hypothesis poses significant challenges. In particular, covariates within and between the groups exhibit strong correlations, as illustrated in Figure~\ref{fig3}. As a result, the estimated signal strength of $\bbeta_m^*$ may be artificially inflated by the influence of nuisance covariate groups, even when the null hypothesis is true. When testing $H_0$, ignoring these correlations may lead to false discoveries.
Moreover, the decorrelated score method is designed to eliminate the impact of nuisance covariates \citep{jrssb2017}.
However, it focuses on scenarios where the dimension of $\bbeta_m^*$ is fixed.
When the dimension of $\bbeta_m^*$ diverges as the sample size increases, the decorrelated score cannot be applied directly due to its null limiting distribution deviating from the normal distribution.

To address these issues, we assume that the $m$-th group of covariates $\bx_m$ are driven by some common latent factors:
	\begin{align} \label{factor}
	\bx_m=\bB_m \vf_m+\bu_m,
	\end{align}
	where $\vf_m \in \bbR^{K_m}$ denotes the vector of random latent factors with $\bbE[\vf_m]=\mathbf 0 $, 
    $\bu_m \in \bbR^{p_m}$ is the vector of random idiosyncratic errors with $ \bbE[\bu_m]=\mathbf 0$ and uncorrelated with $\vf_m$, 
    and $\bB_m \in \bbR^{p_m \times K_m}$ is the loading matrix of $\bx_m$ on the latent factors $\vf_m$. 
    Define $\bbeta=(\bbeta_1^\top,\dots,\bbeta_M^\top)^\top$ and $\bbeta_{-m}=(\bbeta_1^\top,\dots,\bbeta_{m-1}^\top,\bbeta_{m+1}^\top,\dots,\bbeta_M^\top)^\top$.
    Let $\bbeta^*$ and $\bbeta_{-m}^*$ be the true values of $\bbeta$ and $\bbeta_{-m}$, respectively.
    Define $\bx_{-m}=(\bx_1^\top,\dots,\bx_{m-1}^\top,\bx_{m+1}^\top,\dots,\bx_{M}^\top)^\top$.
    
    We perform factor decomposition on the $m$-th group of covariates in model (\ref{cox}):
	\begin{align}\label{2.4}
		\lambda(t|  \mathcal{X})=\lambda_0(t)\exp\left(\bx_{-m}^\top\bbeta_{-m}^*+\vf_m^\top\bgamma_m^*+\bu_m^\top\bbeta_m^*\right),
	\end{align}
	where $\bgamma_m^* = \bB_m^\top\bbeta_m^*$. 
     We call this model as the MOFA-Cox model. The factor model (\ref{factor}) establishes a framework in which the latent factor vector $\vf_m$ captures systematic dependencies, 
     while the idiosyncratic error vector $\bu_m$ behaves as an independent pseudo-covariate.
     It is important to note that for testing $H_0$, the factor decomposition does not need to be uniformly applied across other groups,
     thereby alleviating the influence of idiosyncratic errors from those groups. 
   
    Under the null hypothesis, we have $\boldsymbol{\gamma}_m^*=0$.
    This implies that the testing problem \eqref{2.3} can be reformulated as
	\begin{align}\label{2.5}
		H_0: \boldsymbol{\gamma}_m^*=\mathbf{0} \quad \text { versus } \quad H_a: \boldsymbol{\gamma}_m^* \neq \mathbf{0}.
	\end{align}
    This reformulation utilizes the factor model as a dimension reduction procedure, converting a high-dimensional test for \( \bbeta^*_m \in \mathbb{R}^{p_m} \) into a lower-dimensional test for \( \bgamma_m^* \in \mathbb{R}^{K_m} \).

\setcounter{equation}{0}

\section{Factor-adjusted  test procedure}\label{sec:FADS}
This section introduces our proposed decorrelated score test for assessing a group of covariates. We develop our decorrelated score test in subsection~\ref{sec:decorr}. Factor estimation methods are provided in subsection~\ref{sec:factor}.
\subsection{Factor-adjusted decorrelated score test}\label{sec:decorr}
In this subsection, we develop a Factor-Adjusted Decorrelated Score (FADS) test for (\ref{2.5}). 
 Let $\btheta_m=(\bgamma_m^\top,\bbeta^\top)^\top$ and $\btheta_m^*=(\bgamma_m^{*\top},\bbeta^{*\top})^\top$ be the true value of $\btheta_m$. 
 Define $\bz= (\bx_{1}^\top,\ldots,\bx_{m-1}^\top,\bu_{m}^\top, \bx_{m+1}^\top, \ldots,\bx_{M}^\top)^\top$.
   Suppose that $\{Y_i(t),N_i(t), \delta_i,\bx_{i,l}, l\in[M]\}~(i\in[n])$ are $n$ independent and identically distributed observations of $\{Y(t), N(t), \delta, \bx_{l}, l\in[M]\}$,
and $\{\vf_{i,m}, \bu_{i,m}\}~(i\in[n])$ are independent and identically distributed true latent variables pertaining to the $n$ subjects.We assume that there are no tied observations.
      By the factor model~(\ref{factor}), we have 
      $$\bX_m= \bF_m\bB_m^\top+\bU_m,$$ 
      where $\bX_m=(\bx_{1,m},\dots,\bx_{n,m})^\top, \bF_m=(\vf_{1,m},\dots,\vf_{n,m})^\top$, and $\bU_m=(\bu_{1,m},\cdots,\bu_{n,m})^\top$. 
Next, we fix some notation. For $k=0,~1$ and $2$, we define 
	$$
	\begin{aligned}
		\bPhi_k(t,\btheta_m;\vf_m)&=\frac{1}{n}\sum_{i=1}^n \{(\vf_{i,m}^\top,\bz_i^\top)^\top\}^{\otimes k}Y_i(t)\exp(\vf_{i,m}^\top\bgamma_m+\bz_i^\top\bbeta),\\
        \mbox{and}~~ \bphi_k(t,\btheta_m;\vf_m)&=\bbE\Big[\{(\vf_{m}^\top,\bz^\top)^\top\}^{\otimes k}Y(t)\exp(\vf_{m}^\top\bgamma_m+\bz^\top\bbeta)\Big], \\
	\end{aligned}
	$$
	where for any vector $\balpha,\balpha^{\otimes0}=1,\balpha^{\otimes1}=\balpha, 
        \balpha^{\otimes2}=\balpha\balpha^\top$. Under model \eqref{2.4}, the negative log partial likelihood of $\btheta_m^*$  is defined as
	\begin{align} \label{mle_factor}
		{l(\btheta_m;\vf_m)}=-\frac{1}{n}\sum_{i=1}^n\int_0^\tau(\vf_{i,m}^\top\bgamma_m+\bz_i^\top\bbeta)dN_i(t)+\frac{1}{n}\int_0^\tau\log\bPhi_0(t,\btheta_m;\vf_m)d \bar N(t),
	\end{align}
    where $\bar N(t)=\sum_{i=1}^n N_i(t)$.
Denote $\ell(\btheta_m;\vf_m)$ as the population counterpart of $l(\btheta_m;\vf_m)$. 
Let $\bSigma(\btheta_m;\vf_m)=\nabla_{\btheta_m,\btheta_m}^2\ell(\btheta_m;\vf_m)$, $\bSigma_{\bgamma_{m},\bgamma_{m}}(\btheta_m;\vf_m)=\nabla_{\bgamma_{m},
\bgamma_{m}}^2\ell(\btheta_m;\vf_m)$, $\bSigma_{\bbeta_{-m},\bgamma_m}(\btheta_m;\vf_m)=\nabla_{\bbeta_{-m},\bgamma_m}^2\ell(\btheta_m,\vf_m)$ 
and $\bSigma_{\bbeta_{-m},\bbeta_{-m}}(\btheta_m;\vf_m)=\nabla_{\bbeta_{-m},\bbeta_{-m}}^2\ell(\btheta_m;\vf_m)$. 
Define $\bSigma^*=\bSigma(\btheta_m^*; \vf_m)$. 
Additionally, $\bSigma^*_{\bgamma_m,\bgamma_m}$, ${\bSigma}^*_{\bbeta_{-m},\bbeta_{-m}}$ and $\bSigma^*_{\bbeta_{-m},\bgamma_m}$ is defined similarly. Let  $\bW(\btheta_m;\vf_m) = {\bSigma}^{-1}_{\bbeta_{-m},\bbeta_{-m}}(\btheta_m;\vf_m) \bSigma_{\bbeta_{-m},\bgamma_m}(\btheta_m;\vf_m)$, which can be viewed as a projection matrix of $\nabla_{\gamma_{m}}\ell(\btheta_m;\vf_m)$ onto the linear space of $\nabla_{\bbeta_{-m}}\ell(\btheta_m;\vf_m)$.  Let  $\bW^* = {\bSigma}^{*-1}_{\bbeta_{-m},\bbeta_{-m}} \bSigma_{\bbeta_{-m},\bgamma_m}^*$.

Based on  \eqref{mle_factor}, we define the factor-adjusted decorrelated score function for the \( m \)-th group of covariates as  
\begin{align} \label{score}
	\bS(\btheta_m;\bxi_m) = -\frac{1}{n}\sum_{i=1}^n\int_0^\tau\bigg[\bxi_{i,m} - \frac{\sum_{j=1}^n Y_j(t)\exp(\vf_{j,m}^\top\bgamma_m + \bz_j^\top\bbeta)\bxi_{j,m}}{\sum_{j=1}^n Y_j(t)\exp(\vf_{j,m}^\top\bgamma_m + \bz_j^\top\bbeta)}\bigg] dN_i(t),
\end{align}
where \( \bxi_{i,m} = \vf_{i,m} - {\bW (\btheta_m;\vf_m)}^\top \bx_{i,-m} \) represents the residual component of the \( m \)-th latent factor after linearly projecting out the effects of other covariate groups. 
This projection procedure is analogous to those described in \cite{annals2017} and \cite{jrssb2017}. 
It results in a decorrelated score for the $m$-th group of covariates, effectively eliminating the influence of other covariate groups on $\bx_m$. 
Furthermore, this projection is essential for controlling the variability of higher-order terms when establishing the central limit theorem, as outlined in Theorem \ref{theorem1}.
The construction of $\bS(\btheta_m;\bxi_m)$ relies on the uncorrelatedness between \( \vf_m \) and \( \bu_m \) 
and the identifiability assumption for $\vf_m$ and $\bB_m$; see Assumptions \ref{asym:SubG} and \ref{asym:Iden} in Section \ref{asymptotic}.

   With a slight abuse of notation,  we write $l(\bbeta_{-m},\bgamma_{m};\vf_m)=l(\btheta_{m};\vf_m)$ 
   and $\bS(\bbeta_{-m},\bgamma_m;\bxi_m)=\bS(\btheta_m;\bxi_m)$ when $\bbeta_m=\boldsymbol{0}$.
Let $\hat{\boldsymbol{F}}_m=(\hat{\vf}_{1,m},\dots,\hat{\vf}_{i,m})^\top$ be an estimate of the latent factor matrix $\boldsymbol{F}_m$.
The estimation of ${\boldsymbol{F}}_m$ is defined in subsection~\ref{sec:factor}.
Under the null hypothesis, we can estimate $\bS(\bbeta_{-m},\bgamma_m;\bxi_m)$ by
	\[
	\bS(\hat{\bbeta}_{-m},\mathbf{0};\hat{\bxi}_m)=-\frac{1} 
        {n}\sum_{i=1}^n\int_0^\tau\left[\hat{\bxi}_{i,m}-\frac{\sum_{j=1}^nY_j(t)\exp(\bx_{j,-m}^\top\hat\bbeta_{-m})\hat{\bxi}_{j,m}}{\sum_{j=1}^nY_j(t)\exp(\bx_{j,-m}^\top\hat\bbeta_{-m})}\right]dN_i(t),
	\]
 where $\hat{\bxi}_{i,m}=\hat\vf_{i,m}-\hat\bW^\top \bx_{i,-m}$ and $\hat{\bbeta}_{-m}$ denotes an estimate of $\bbeta_{-m}^*\in \bbR^{p-p_m}$. 
 Inspired by \cite{annals2017} and \cite{jrssb2017}, $\hat\bbeta_{-m}$ and the $k$-th column of $\hat\bW \in \bbR^{(p-p_m)\times K_m}$ can be obtained by solving the following optimization problems:
	\begin{align}
		(\hat{\bbeta}_{-m},\hat{\bgamma}_{m})=\text{argmin}_{({\bbeta}_{-m},{\bgamma}_{m})}~~{l}(\bbeta_{-m},\bgamma_{m};\hat{\vf}_m) + \lambda_1\|\bbeta_{-m}\|_1,\label{beta_gamma_hat}
	\end{align}
    and 
    \begin{align}
        &{\hat\bw}_k=\text{argmin} ~~\|{\bw}_k\|_1,  \notag \\
		\text{subject~~to}~~&\left\|\nabla^2_{\bbeta_{-m},\bgamma_m} {l}(\hat\bbeta_{-m},\hat\bgamma_m;\hat{\vf}_m)- \nabla^2_{\bbeta_{-m},\bbeta_{-m}} {l}(\hat\bbeta_{-m},\hat\bgamma_m;\hat{\vf}_m){\bw}_k\right\|_\infty\le\lambda_2. \label{w_hat}
	\end{align}
 Here $\lambda_1$ and $\lambda_2$ denote the tuning parameters and can be selected using standard cross-validation procedures.

	
	Now, we evaluate the variance of the decorrelated score function, denoted as $\bSigma^*_{\bgamma_m|\bbeta_{-m}}$. A direct calculation gives
	$$
	\begin{aligned}
		\bSigma^*_{\bgamma_m|\bbeta_{-m}}= \bSigma^*_{\bgamma_m,\bgamma_m}-\bSigma^*_{\bgamma_m,\bbeta_{-m}}{\bSigma}^{*-1}_{\bbeta_{-m},\bbeta_{-m}}\bSigma^*_{\bbeta_{-m},\bgamma_m}\in \bbR^{K_m\times K_m},
	\end{aligned}
	$$
	which can be estimated by
	$$
	\hat\bSigma_{\bgamma_m|\bbeta_{-m}} = \nabla^2_{\bgamma_m,\bgamma_m} {l} 
       (\hat\bbeta_{-m},\hat\bgamma_m;\hat{\vf}_m)-\hat\bW^\top \nabla^2_{\bbeta_{-m},\bgamma_m} {l}(\hat\bbeta_{-m},\hat\bgamma_m;\hat{\vf}_m).
	$$
	Define
	\[
	\bT_n = \sqrt{n}\hat\bSigma_{\bgamma_m|\bbeta_{-m}}^{-1/2}\bS(\hat{\bbeta}_{- 
        m},\mathbf{0};\hat{\bxi}_m).
	\]
	In Section~\ref{asymptotic}, we show that under the null hypothesis, the asymptotic distribution of $\bT_n$ is $N(\mathbf 0,\bI_{K_m})$. Then our FADS test statistic is given by $\|\bT_n\|_2^2$. Given a significance level $\alpha\in(0,1)$, we reject the null hypothesis if 
        $$\|\bT_n\|_2^2\ge \chi^2_{K_m}(\alpha),$$
     where $\chi^2_{K_m}(\alpha)$ is the $\alpha$-th upper-quantile of a chi-squared distribution with $K_m$ degrees of freedom.

 We offer a few remarks regarding our factor-adjusted decorrelated score function and its comparison with \cite{jrssb2017}. 
First, \cite{jrssb2017} directly incorporates the covariate $\bx_m$ into their score test, and we utilize the latent factors of $\bx_m$ in our decorrelated score test. 
Because these latent factors are unobserved and must be estimated from the data, this complicates the analysis of the asymptotic properties of the score test; see Theorem \ref{theorem1} below. 
Second, we include $\bgamma_m$ in \eqref{beta_gamma_hat} to enhance the estimation accuracy of $\hat\bbeta_{-m}$, 
particularly under the alternative hypothesis. This improves the test's power.
Finally, to achieve a consistent estimation of $\bW^*$, we must substitute the latent factors with their corresponding estimators and then solve a non-typical Dantzig problem \eqref{w_hat}.
This further renders our method non-trivial in theoretical analysis compared to \cite{jrssb2017}.

\subsection{Factor estimation}\label{sec:factor} \label{factor_estimation}
        In the context of the MOFA-Cox model, the latent factor vectors $\vf_m$ and $\bu_m$ must be estimated from the data, 
        given that only the covariates $\bx_m$ are directly observed. 
        To ensure the identifiability of $\bB_m$ and $\vf_m$, we adopt the usual assumptions as in \cite{fan2013}; see Assumptions \ref{asym:SubG} and \ref{asym:Iden} below for more details.
     Furthermore, as indicated in \cite{fan2024}, $K_m$ is associated with the number of spiked eigenvalues of $\bX_m\bX_m^\top$, which is typically small. 
     Thus, we treat $K_m$ as a fixed constant. 
     We consider the constrained least squares estimator for $(\bB_m,\bF_m)$:
        $$
        \begin{aligned}
        &(\hat\bB_m,\hat\bF_m) = \argmin_{\bB_m\in\mathbb R^{p_m\times K_m},\bF_m\in\mathbb R^{n\times K_m}} \|\bX_m-\bF_m\bB_m^\top\|_F^2, \\
        \text{subject~~to}~~&\quad \frac{1}{n}\bF_m^\top\bF_m=\bI_{K_m}\quad \text{and}\quad \bB_m^\top\bB_m~~\text{is diagonal}.
        \end{aligned}
        $$
        Direct calculations yield that $n^{-1/2}\hat\bF_m$ is equal to the eigenvectors corresponding to the largest $K_m$ eigenvalues of $\bX_m\bX_m^\top$,
        $\hat\bB_m=n^{-1}\bX_m^\top\hat\bF_m$ and $\hat\bU_m=(\bI_n-n^{-1}\hat\bF_m\hat\bF_m^\top)\bX_m$; see \cite{fan2013} for more details.
        
        Since the number of latent factors $K_m$ is unknown, it must be determined using the data. 
        Here, we use the ratio method \citep{luo2009,lam2012,ahn2013}, which performs well in our simulation studies and real data analysis.
        Specifically, let $\lambda_k(\bX_m\bX_m^\top)$ denote the $k$-th largest eigenvalue of $\bX_m\bX_m^\top$.
        We estimate the number of factors using
        $$
        \hat K_m = \argmax_{k\le\bar K_m}\frac{\lambda_k(\bX_m\bX_m^\top)}{\lambda_{k+1}(\bX_m\bX_m^\top)},
        $$
        where $\bar{K}_m$ is a prescribed integer. 

\setcounter{equation}{0}
       
        \section{Asymptotic properties} \label{asymptotic}

        In this section, we establish the limiting distributions of the FADS test statistic under the null hypothesis. 
        We also study the local power property under an alternative hypothesis. 
        To proceed, we consider the following regularity assumptions. 
        
\begin{asmp}[Sub-Gaussian condition]\label{asym:SubG} {
    For $m\in[M]$, $\left\{\vf_{i,m}\right\}_{i=1}^n$ and $\left\{\bu_{i,m}\right\}_{i=1}^n$ are i.i.d. uncorrelated sub-Gaussian random vectors with zero mean. 
    Moreover, for all $k\in[K_m]$, $\bx_{i,-m}^\top\bw_k^*$ are i.i.d. sub-Gaussian random vectors with zero mean, where $\bw_k^*$ denotes the $k$-th column of $\bW^*$.
    }
\end{asmp}
     \begin{asmp}[Identifiability]\label{asym:Iden} {
       The factor $\vf_m$ satisfies $\operatorname{var}(\vf_m)=\bI_{K_m}$. In addition, $\bB_m^{\top} \bB_m$ is a diagonal matrix.
        }
\end{asmp}
\begin{asmp}
[Pervasiveness]\label{asym:Per}{
        The eigenvalues of $\bB_m^{\top}\bB_m/p_m$ are bounded away from 0 and $\infty$ as $p_m\to\infty$,
       and $\|\bSigma_{u_m}\|_2$ is bounded. Here, $\bSigma_{u_m}$ denotes the covariance matrix of $\bu_m$.
        }
\end{asmp}

Assumption~\ref{asym:SubG} is typical and commonly adopted in high-dimensional literature \citep{fan2013,li2021integrative,fan2024}. 
Assumption~\ref{asym:Iden} is used to ensure the identifiabilities of $\vf_m$ and $\bB_m$ \citep{bai2003,fan2013}. 
Under Assumption~\ref{asym:Iden}, the covariance matrix of $\bx_m$ is $\bSigma_m=\bB_m\bB_m^\top+\bSigma_{\bu_m}$. 
Thus, we can apply Principal Component Analysis (PCA) to discover the latent factors, the loading matrix and the idiosyncratic errors.
The pervasiveness assumption (Assumption~\ref{asym:Per}) is widely accepted in the application of PCA, 
as it ensures that the influence of noise on $\bx_m$ is negligible compared to the latent factor $\vf_{m}$ \citep{fan2013,li2018,fan2022}.

Let $\hat\bSigma_m=n^{-1}\bX_m^\top\bX_m$ be the sample covariance matrix of the $m$-th covariate group.
Let $\hat\lambda_{k}$ denote the $k$-th largest eigenvalue of $\hat\bSigma_m$, and let $\hat{\boldsymbol\eta}_{k}$ represent the corresponding orthonormalized eigenvector.
Define $\hat\bLambda_m=\text{diag}(\hat\lambda_1,\dots,\hat\lambda_{K_m})$ and $\hat\bGamma_m=(\hat{\boldsymbol\eta}_1,\dots,\hat{\boldsymbol\eta}_{K_m})$.

       \begin{asmp}[Loadings and initial pilot estimators]\label{asym:lod}{
        For $m\in[M]$, $\|\bB_m\|_\infty$ is bounded. In addition, $\hat\bSigma_m,\hat\bLambda_m$ and $\hat\bGamma_m$ satisfy $\|\hat\bSigma_m-\bSigma_m\|_\infty=O_p\{\sqrt{\log (p_m)/n}\},\|(\hat\bLambda_m-\bLambda_m)\hat\bLambda_m^{-1}\|_\infty=O_p\{\sqrt{\log (p_m)/n}\}$ and $\|\hat\bGamma_m-\bGamma_m\|_\infty=O_p\{\sqrt{\log (p_m)/(np_m)}\}$.
        }
\end{asmp}

         Assumption~\ref{asym:lod} is widely adopted in various relevant scenarios, such as the sample covariance matrix under sub-Gaussian distributions \citep{fan2013}. It also holds for the marginal and spatial Kendall's tau estimators \citep{fan2018} as well as the elementwise adaptive Huber estimator \citep{fan2019}. 
        

Define $\mathcal O=\text{supp}(\bbeta_{-m}^*),\mathcal O^c=[p-p_{m}]/\mathcal O$ and $s_{-m}^*=\left|\mathcal O\right|$. Following \cite{lasso2013}, we define the cone that for some constant $\xi>1$,
        $$
        \mathcal C(\xi;\mathcal O)=\left\{\balpha\in\mathbb R^{p_{-m}}:\Vert\balpha_{\mathcal O^c}\Vert_1\le\xi\Vert\balpha_\mathcal O\Vert_1\right\},
        $$
        and the compatibility factor 
        $$
        \kappa(\xi;\mathcal O)=\inf_{\bzero\neq\balpha\in\mathcal C(\xi;\mathcal O)}\frac{s_{-m}^{*1/2}\{\balpha^\top\nabla^2_{\bbeta_{-m},\bbeta_{-m}}l(\bbeta_{-m}^*,\bgamma_m^*;\vf_m)\balpha\}^{1/2}}{\Vert\balpha_\mathcal O\Vert_1}.
        $$
        
                \begin{asmp}[Compatibility factor]\label{asym:Com} {
        There exists a positive constant $C_\kappa$ such that $\kappa(\xi;\mathcal O)\ge C_\kappa$ holds with probability one.
        } 
        \end{asmp}
        
        
        Assumption~\ref{asym:Com} is known as the compatibility factor condition, which essentially bounds the minimal eigenvalue of $\nabla^2_{\bbeta_{-m},\bbeta_{-m}}l(\bbeta_{-m}^*,\bgamma_m^*;\vf_m)$ from below for those directions within the cone $\mathcal C(\xi;\mathcal O)$. 
        This assumption is adopted to establish the convergence rate of $\hat\bbeta_{-m}$ and $\hat\bW$,
        and its validity has been verified in Theorem 4.1 of \cite{lasso2013}.

        \subsection{Asymptotic null distribution}\label{sec:4.1}

        In this subsection, we establish the limiting null distribution of the FADS test statistic. 
         In the following theoretical analysis, we treat $K_m$ as known. 
         All theoretical results remain valid, as $\hat K_m$ is a consistent estimate of $K_m$ \citep{fan2022,fan2024}.
        A key step in our theoretical analysis is to ensure the consistency of the estimators $\hat\bbeta_{-m}$ and $\hat\bW$ in (\ref{beta_gamma_hat}) and (\ref{w_hat}). 
         Under Assumption~\ref{asym:Com}, along with the condition that covariates are uniformly bounded, \cite{lasso2013} established the convergence rate of the Lasso estimator $\hat\bbeta$ under the $l_1$ norm. However, their analysis does not account for the influence of latent factor estimation on the accuracy of $\hat\bbeta$, which is pertinent in our context. 
         Therefore, we require to develop new methods to establish the convergence rate of $\hat\bbeta$. 

         Define    
         $$
         \begin{aligned}
            C_{n,p_m,p-p_m} &=\sqrt\frac{\log(p_m)\log(n)}{n}+\sqrt{\frac{\log(n)}{p_m}}+\sqrt{\log[n(p-p_m)]}\sqrt{\frac{\log(p-p_m)}{n}}, \\
            \text{and}~~C_{n,p_m,p-p_m}^\prime &=\left[\sqrt\frac{\log(p_m)\log(n)}{n}+\sqrt{\frac{\log(n)}{p_m}}+1\right]\sqrt{\frac{\log(p-p_m)}{n}}.
        \end{aligned}
        $$
  

        \begin{lemma}\label{lemm2}
        Suppose that Assumptions~\ref{asym:SubG}-\ref{asym:Com} hold. If $\lambda_1\asymp C_{n,p_m,p-p_{m}}$ and $\lambda_2\asymp C_{n,p_m,p-p_{m}}^\prime$,
        then we have
        $$
        \big\|\hat{\bbeta}_{-m}-\bbeta_{-m}^*\big\|_1=O_p\left(s_{-m}^*\lambda_1\right) \quad \text{and} \quad
        \big\|\hat{\bw}_k-\bw_k^*\big\|_1=O_p\left(s_k^*\lambda_2\right),
        $$
         where $\bw_k^*$ denotes the $k$-th column of $\bW^*$ and $s_k^*=\left|\text{supp}(\bw_k^*)\right|$.
        \end{lemma}

        Lemma \ref{lemm2} is a byproduct of our research, which may be of interest when analyzing high-dimensional survival data with factor adjustment.
        It provides an upper bound for the LASSO estimator of $\bbeta_{-m}^*$ and $\bw_k^*$.
        For illustration, we take $\hat\bbeta_{-m}$ as an example. The statements for $\hat\bw_k$ can be obtained in a similar way.
        By the definition of $C_{n,p_m,p-p_{m}}$, the upper bound of $\|\hat{\bbeta}_{-m}-\bbeta_{-m}^*\|_1$ becomes
        \begin{align}\label{Lem:eq1}
            s_{-m}^*\bigg(\sqrt\frac{\log(p_m)\log(n)}{n}+\sqrt{\frac{\log(n)}{p_m}}+\kappa_{n,p}\sqrt{\frac{\log(p-p_m)}{n}}\bigg)
        \end{align}
        with $\kappa_{n,p}=\sqrt{\log[n(p-p_m)]}.$ The first two terms of \eqref{Lem:eq1}  arise from the estimation errors for the latent factors ${\vf}_{i,m}$. 
        In \cite{jrssb2017}, a similar upper bound is required for theoretical analysis:
        \begin{align*}
           \big\|\hat{\bbeta}_{-m}-\bbeta_{-m}^*\big\|_1=O_p\Big(s_{-m}^* \sqrt{\log(p-p_m)/n}\Big),
        \end{align*}
        which corresponds to the third term of \eqref{Lem:eq1}.
        While they assume that the covariates are bounded  by a universal constant, we consider a sub-Gaussian assumption for covariates.
        Consequently, $\kappa_{n,p}$ arises from the application of the maximal inequality for sub-Gaussian variables \citep{concentration}; see the Supplementary Material for details.
        If the covaraites are bounded, then $\kappa_{n,p}$ in \eqref{Lem:eq1} can also be a universal constant as in \cite{jrssb2017}.
        
        The following theorem presents the asymptotic normality of the FADS test statistic under the null hypothesis.

       \begin{theorem} \label{theorem1}
        Suppose that Assumptions \ref{asym:SubG}-\ref{asym:Com} hold. 
        If $\lambda_1\asymp C_{n,p_m,p-p_{m}}$, $\lambda_2\asymp C_{n,p_m,p-p_{m}}^\prime$,
        $n^{1/2}s^*_{-m}C_{n,p_m,p-p_m}C_{n,p_m,p-p_m}^\prime=o(1)$, and $s_w^*C_{n,p_m,p-p_m}^\prime\sqrt{\log(p-p_m)}=o(1)$,
        then under the null hypothesis $\bbeta_m^*=\mathbf{0}$, we have
        $$
        \bT_n\stackrel{{d}}{\longrightarrow} N(\mathbf{0},\bI_{K_m}),
        $$
        where $\stackrel{{d}}{\longrightarrow}$ denotes convergence in distribution, $s_w^*=\max_{k\in[K_m]} s_k^*$ and $s_k^*$ are defined in Lemma \ref{lemm2}. 
           \end{theorem}
    
       Under Theorem \ref{theorem1}, the FADS test achieves a significance level of $\alpha$ under the null hypothesis $H_0$.
       Next, we consider the assumptions of Theorem \ref{theorem1}. For comparison and illustration, we assume that the covariates are bounded and focus on analyzing the assumptions regarding $\lambda_1$ and $n^{1/2}s^*_{-m}C_{n,p_m,p-p_m}C_{n,p_m,p-p_m}^\prime=o(1)$.
       \begin{case}[\(\log(p_m)\asymp n^{c_0}\) with $c_0\in [0,1)$ and $p_m/p\rightarrow 0$ as $n\rightarrow\infty$]\label{caseI}
      This case corresponds to the setting where the dimension of the covariates of interest is large, yet remains negligible compared to that of the nuisance groups.
       In this case, we obtain \(\lambda_1 \asymp \sqrt{\log(p-p_{m})/n} \), $\lambda_2\asymp \sqrt{\log(p-p_{m})/n}$,
       \begin{align*}
       &n^{1/2}s^*_{-m}C_{n,p_m,p-p_m}C_{n,p_m,p-p_m}^\prime=O\big(s^*_{-m}n^{-1/2}\log(p- p_m)\big)=o(1),\\
       \text{and}~~~~&s_w^*C_{n,p_m,p-p_m}^\prime\sqrt{\log(p-p_m)}=O\big(s^*_wn^{-1/2}\log(p-p_m)\big)=o(1).
       \end{align*}
       These assumptions align with those in Theorem 1 of \cite{jrssb2017}. 
       Furthermore, the order of the upper bounds of $\hat\bbeta_{-m}$ and $\hat\bW$ achieves the oracle, 
       as obtained when the latent factors are known. 
       This implies that, in the case of the exponential rate with $p_m/p\rightarrow 0$,
       there is no additional cost to infer $\bbeta_m^*$ when using our method with estimated factors. 
       \end{case}
      
     \begin{case} [\(\log(p_m)\asymp n^{c_0}\) with $c_0\in [0,1)$ and $p_m/p\rightarrow c_1\in(0,1)$ as $n\rightarrow\infty$]\label{caseII}
      This case corresponds to the setting where the dimension of the covariates of interest is large, yet remains negligible compared to that of the nuisance groups.
       In this case, we obtain \(\lambda_1 \asymp \sqrt{\log(p_m)\log(n)/n} \) and
       \begin{align*}
       &n^{1/2}s^*_{-m}C_{n,p_m,p-p_m}C_{n,p_m,p-p_m}^\prime=O\big(s^*_{-m}n^{-1/2}\log(p-p_m)\sqrt{\log(n)}\big)=o(1),\\
       \text{and}~~~~&s_w^*C_{n,p_m,p-p_m}^\prime\sqrt{\log(p-p_m)}=O\big(s^*_wn^{-1/2}\log(p-p_m)\big)=o(1).
       \end{align*}
       The assumptions for $\lambda_2$ and $s_w^*C_{n,p_m,p-p_m}^\prime\sqrt{\log(p-p_m)}$ are the same as that in Case~\ref{caseI}.
       Therefore, the assumptions remain aligned with those of Theorem 1 of \cite{jrssb2017}, ignoring a logarithmic term.
\end{case}

\begin{case}[\(\log(p_m)\asymp n^{c_0}\) with $c_0\in [0,1)$ and $p_m/p\rightarrow 1$ as $n\rightarrow\infty$]\label{caseIII}
     In this case, the dimension of the covariates in the nuisance groups is negligible compared to that of the covariates in the group of interest.
       We then obtain
       \begin{align*}
        n^{1/2}s^*_{-m}C_{n,p_m,p-p_m}C_{n,p_m,p-p_m}^\prime=O\big(s^*_{-m}n^{-1/2}\log(p_m)\sqrt{\log(n)}\big)=o(1).
       \end{align*} 
       The assumptions for \(\lambda_1\), $\lambda_2$ and $s_w^*C_{n,p_m,p-p_m}^\prime\sqrt{\log(p-p_m)}$ are the same as in Case~\ref{caseII}.
       In this case , the estimation error of $\hat\bbeta_{-m}$ is dominated by that of the estimated factors.
       Although the convergence rate does not attain the oracle rate, Theorem~\ref{theorem1} still holds.
       Based on Cases~\ref{caseI}-\ref{caseIII}, our method is particularly well-suited for testing the significance of a group whose covariate dimension grows at an exponential rate with respect to the sample size. 
\end{case}
     \begin{case}[$p_m=O(n^{c_0})$ with $c_0>0$ and $\log(p-p_{m})=O(n^{c_1})$ with $c_1\in[0,1)$]\label{caseIV}
       If $c_1\ge 1-c_0$, 
       then the assumptions are the same as in Case~\ref{caseI}. 
       This implies that the estimation error caused by $\hat\vf_{i,m}$ can be negligible when inferring $\bbeta_m^*$. 
       If $c_1\in[0,1-c_0)$ with $c_0\in (0,1)$, the estimation error caused by $\hat\vf_{i,m}$ is non-negligible.
       However, Theorem \ref{theorem1} remains valid with $\lambda_1\asymp \sqrt{\log(n)/p_m}$ and
       $$
       n^{1/2}s^*_{-m}C_{n,p_m,p-p_m}C_{n,p_m,p-p_m}^\prime=O\big(s^*_{-m}p_m^{-1/2}\sqrt{\log(n)}\big)=o(1).
       $$
      Therefore, Case~\ref{caseIV} shows that our method is applicable when the group of interest has moderate dimensionality, whereas the nuisance groups are high-dimensional.  
     \end{case}

        \subsection{Power analysis}

      In this subsection, we study the power of the FADS test under the local alternative $H_{a_n}:\, \bbeta_m^*=\boldsymbol b_{m_n}$, 
      where $\boldsymbol b_{m_n}$ denotes a vector sequence approaching zero as $n\to\infty$. 
      We consider the following parameter space for the local alternative: 
        $$
        \mathcal{N}=\left\{\bbeta^*\neq \mathbf 0_p:\bbeta^*_m=\boldsymbol{b}_{m_n}~~\text{and}~~\left|\operatorname{supp}\left(\bbeta_{-m}^*\right)\right|=s_{-m}^*~\text {with }~s_{-m}^* \ll n\right\}.
        $$
        Define $\boldsymbol c_{m_n}=\bB_m^\top\boldsymbol{b}_{m_n}$. The next theorem establishes the limiting distribution of $\|\bT_n\|_2^2$ under the local alternative.
        
      \begin{theorem} \label{theorem2}
           Suppose that the conditions of Theorem~\ref{theorem1} hold. If $\left\|\boldsymbol{b}_{m_n} \right\|_2=o(1/\sqrt{\log n}) $ and $\left\|\boldsymbol{c}_{m_n} \right\|_2=o(1/\sqrt{\log n})$, 
           then we have
        $$
        \sup _{x>0}\left|\mathbb P\left(\|\bT_n\|_2^2 \leq x\right)-\mathbb P\left\{\chi^2\left(K_m, h_{m_n}\right) \leq x\right\}\right| \rightarrow 0
        $$
        uniformly for all $\bbeta^*\in\mathcal{N}$,
        where $h_{m_n}=n \boldsymbol{c}_{m_n}^{\top} \bSigma^*_{\boldsymbol{\gamma}_m \mid \boldsymbol{\beta}_{-m}} \boldsymbol{c}_{m_n}$ 
        and $\chi^2(K_m, h_{m_n})$ is a non-central chi-squared random variable with $K_m$ degrees of freedom and non-centrality parameter $h_{m_n}$.
\end{theorem}

Note that $h_{m_n}\asymp n\|\boldsymbol{c}_{m_n}\|_2^2$. 
Theorem \ref{theorem2} indicates that $\|\boldsymbol{c}_{m_n}\|_2$ essentially controls the power of the FADS test. 
For illustration, we set $\|\boldsymbol{c}_{m_n}\|_2\asymp n^{-l_m}$ with 
certain $l_m > 0$. 
The power then exhibits transitional behavior based on the value of $l_m$, as provided in the following corollary.
    
      \begin{coro}\label{coro1} 
        Let $h=\lim_{n\to\infty}n \boldsymbol{c}_{m_n}^{\top} \bSigma^*_{\boldsymbol{\gamma}_m \mid \boldsymbol{\beta}_{-m}} \boldsymbol{c}_{m_n}$.
        Suppose that the conditions of Theorem~\ref{theorem2} hold. Then we have
        \begin{description}
        \itemsep=-\parsep
        \itemindent=-1cm
        \item (i) $\lim_{n\to\infty}\sup_{\bbeta^*\in\mathcal N}\sup_{x>0} |\mathbb P(\|\bT_n\|_2^2\le x)-\mathbb P\left\{\chi^2(K_m,0)\le x\right\}|\to 0$ if $l_m>1/2$.
        \item (ii) $\lim_{n\to\infty}\sup_{\bbeta^*\in\mathcal N}\sup_{x>0} |\mathbb P(\|\bT_n\|_2^2\le x)-\mathbb P\left\{\chi^2(K_m,h)\le x\right\}|\to 0$ if $l_m=1/2$.
        \item (iii) For any $x>0$, $\liminf_{n\to\infty}\sup_{\bbeta^*\in\mathcal N}\mathbb P(\|\bT_n\|_2^2>x)=1$ if $l_m<1/2$.
        \end{description}
\end{coro}

The first part of Corollary~\ref{coro1} indicates that if $\|\boldsymbol{c}_{m_n}\|_2$ is of a smaller order of $n^{-1/2}$, 
the power diminishes to the significance level  $\alpha$. In this case, the FADS test can not distinguish the null hypothesis from the local alternatives. 
Additionally, the power is determined by a non-central $\chi^2$-distribution when $l_m=1/2$.
If $\|\boldsymbol{c}_{m_n}\|_2$ has a higher order of $n^{-1/2}$, the power converges to 1, 
which implies that the proposed test is consistent.
This phenomenon is also observed in \cite{li2021integrative} in the context of linear regression models. 


Theorem \ref{theorem2} and Corollary \ref{coro1} demonstrate that the power depends on the relationship between the loading matrix $\bB_m$ 
and the location $\boldsymbol{b}_{m_n}$ of the alternative hypothesis, i.e., $\boldsymbol{c}_{m_n}=\bB_m^\top\boldsymbol{b}_{m_n}$. 
To gain a clearer insight into the power behavior, 
we consider a special case for $\bB_m$ and $\boldsymbol{b}_{m_n}$. 
Let $K_m=1$, $\bB_m=(c_1,\dots,c_{p_m})^\top$ and $\boldsymbol{b}_{m_n}=(d_{n,1},\dots,d_{n,L_n},0,\dots,0)^\top$, where $L_n$ is a nonnegative integer depending on $n$.  
Under these settings, we have $\boldsymbol{c}_{m_n}=\sum_{j=1}^{L_n}c_{j}d_{n,j}$.
For $j\in[L_n]$, let $d_{n,j}\asymp n^{-\eta_1}$ with $L_n\asymp n^{\eta_2}$. 
Here, $\eta_1$ and $\eta_2$ are positive constants.
Suppose that $\|\bB_m\|_\infty$ is bounded as shown in Assumption~\ref{asym:lod}.
Then we have $\boldsymbol{c}_{m_n}\asymp n^{\eta_{2}-\eta_1}$.
By Corollary~\ref{coro1}, the power of the FADS test converges to 1 {when $0<\eta_1-\eta_2<1/2$.}
If we further set $\eta_1>1/2$, then the power vanishes when performing a debiased or decorrelated test on each individual element of $\bbeta_m^*$ \citep{jrssb2017}.
In this case, the power is improved using the factor-adjusted test.
However, when $\bB_m$ and $\boldsymbol{b}_{m_n}$ are orthogonal, the FADS test may exhibit less power to detect the alternative with $\boldsymbol{c}_{m_n}=0$,
regardless of the magnitude of $\boldsymbol{b}_{m_n}$.

We conclude this section by discussing the differences and challenges in theoretical analysis compared to existing methods 
(e.g., \citealp{jrssb2017}, \citealp{li2021integrative}, \citealp{liWang}).
First, although \cite{jrssb2017} considered a decorrelated score test for the Cox model, 
the dimension of the parameters being tested is fixed.
In contrast, we allow the dimension of $\bbeta_m^*$ to grow with $n$, while fixing the dimension of $\bgamma_m^*$. 
Furthermore, since the factors cannot be directly observed, they must be estimated from the data.
This significantly increases the uncertainty of the score function and the Hessian matrix when substituting the latent factors with their estimates,
thereby complicating the analysis of uncertainty for our test.
Lastly, unlike \cite{li2021integrative} and \cite{liWang}, the outcomes are not fully observed, and our score function lacks the characteristic of i.i.d. samples.
Consequently, we must employ concentration inequalities and maximal inequalities to control estimation errors within the framework of counting process theory.

\section{Simulation studies} \label{simulation}

In this section, we evaluate the finite sample performance of the proposed method. 
For comparison, we consider the method presented in \cite{jrssb2017}. 
We set the number of groups to $M=2$ and define the number of covariates in each group as $p_1=p_2=p/2$. 
Under the MOFA-Cox model \eqref{cox}, the hazard function takes the form:
$$
\lambda(t|\bx_1,\bx_2)=\lambda_0(t)\exp(\bx_1^\top\bbeta_1^*+\bx_2^\top\bbeta_2^*).
$$
We set the baseline function $\lambda_0(t)=1$, and  consider the following two cases for the covariate groups $\bx_1$ and $\bx_2$. 

\noindent{\bf Case 1}: Let $K_1=K_2=2$ and $\bX_m=\bF_m\bB_m^\top+\bU_m$ for $m=1$ and $2$. 
The entries of $\bB_m$ are generated from the uniform distribution $U(-1,1)$, 
and each row of $\bU_m$ follows a multivariate normal distribution $N(0,\bSigma_{u_m})$, where $\bSigma_{u_m}=(0.5^{|i-j|})_{1\le i,j\le p_m}$. 
For the $m$-th group, the latent factors $\vf_{i,m}$ are generated from the standard multivariate normal distribution, i.e. $\vf_{i,m}\sim N(0,\bI_{K_m})$~$(i=1,\dots,n)$.
In this case, $\bx_m$ follows a factor model setup. 

\noindent{\bf Case 2}: For each group, $\bx_m\sim N(0,\bSigma_m)$ with $\bSigma_m=(0.5^{|i-j|})_{1\le i,j\le p_m}$. 
In this case, although \( \bx_m \) deviates from the assumptions of the factor model \eqref{factor}, 
the covariance matrix exhibits a spiked eigenvalue structure.
Thus, a factor-based approximation remains theoretically justifiable and potentially effective for dimensionality reduction.
    
We focus on testing whether $\bx_1$ has significant effects on the hazard, that is, $H_0: \bbeta_1^*=0$.
Let $\beta_{m_j}^*$ denote the $j$-th element of $\bbeta_m^*$.
We consider two types of alternatives:
a sparse alternative, i.e., $H_{a1}:\beta_{1,1}^*=\cdots=\beta_{1,5}^*=b_0$ and $\beta_{1,6}^*=\cdots=\beta_{1,{p_1}}^*=0$
and a dense alternative, i.e., $H_{a2}:\beta_{1,1}^*=\cdots=\beta_{1,{p_1}}^*=b_0$. 
We set $b_0\in\{0.05, 0.1, \dots, 0.5\}$ for the sparse alternative
and set $b_0\in\{0.02, 0.04, \dots, 0.2\}$ for the dense alternative. 
In addition, we set $\beta_{2,1}^*=\beta_{2,2}^*=1$ and $\beta_{2,3}^*=\cdots=\beta_{2,{p_2}}^*=0$. 
The censoring time $C$ follows a uniform distribution $U(0,c)$, where $c$ is chosen such that the censoring rate is about $40\%$. 
We set $(n,p)=(200,600)$ and $(n,p)=(200,1000)$.
The nominal level $\alpha$ is chosen as $0.05$.
Each simulation is repeated 500 times to calculate the empirical size and power.

\begin{figure}[thp] 
    \centering
    \begin{subfigure}[b]{0.45\textwidth}
        \centering
        \includegraphics[width=\textwidth]{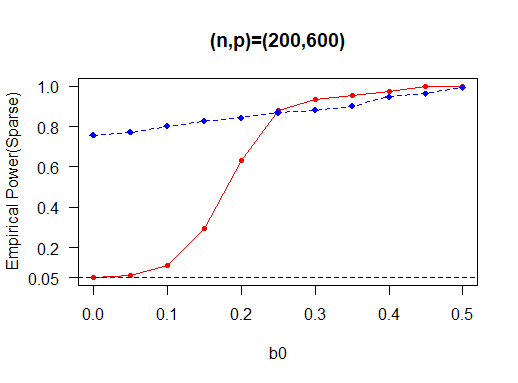}
    \end{subfigure}  
    \hspace{0.05\textwidth}
    \begin{subfigure}[b]{0.45\textwidth}
        \centering
        \includegraphics[width=\textwidth]{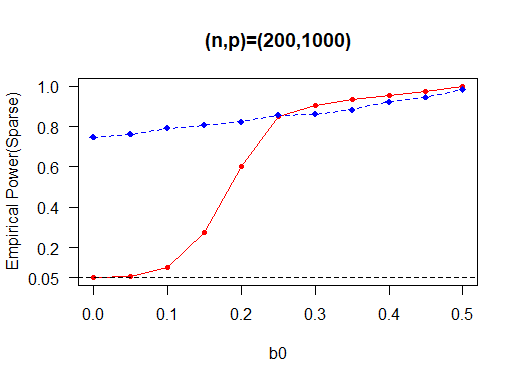}
    \end{subfigure}

    \vspace{0.5cm} 

    \begin{subfigure}[b]{0.45\textwidth}
        \centering
        \includegraphics[width=\textwidth]{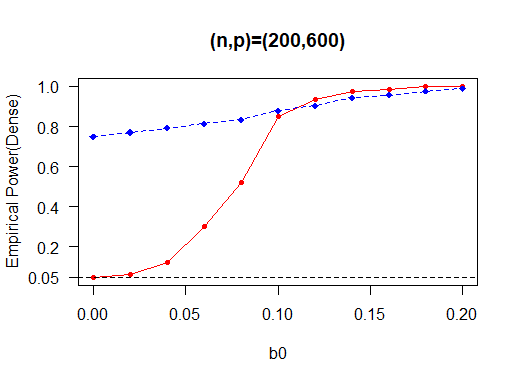}
    \end{subfigure}
    \hspace{0.05\textwidth}
    \begin{subfigure}[b]{0.45\textwidth}
        \centering
        \includegraphics[width=\textwidth]{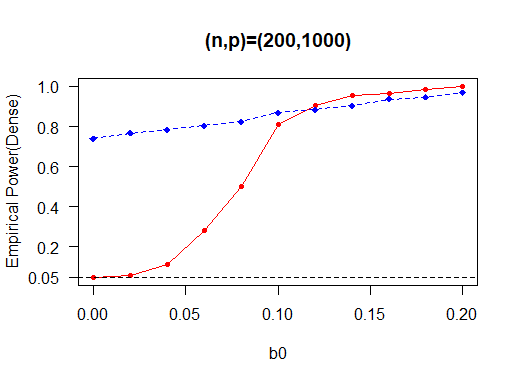}
    \end{subfigure}

    \caption{Empirical sizes and powers for Case 1: the FADS test (solid line) and the test of \cite{jrssb2017} (dashed line).}
    \label{case1}
    \end{figure}

\begin{figure}[thp] 
    \centering
    \begin{subfigure}[b]{0.45\textwidth}
        \centering
        \includegraphics[width=\textwidth]{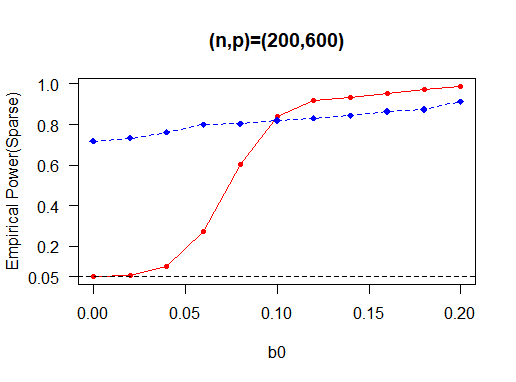}
    \end{subfigure}  
    \hspace{0.05\textwidth}
    \begin{subfigure}[b]{0.45\textwidth}
        \centering
        \includegraphics[width=\textwidth]{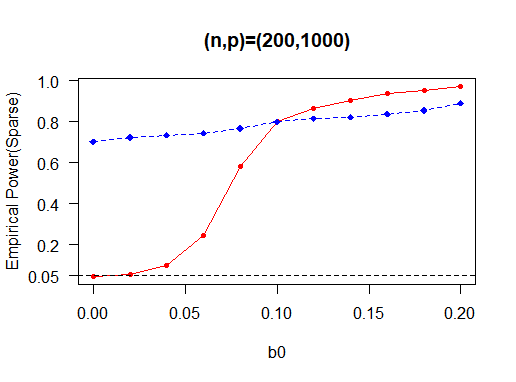}
    \end{subfigure}

    \vspace{0.5cm} 

    \begin{subfigure}[b]{0.45\textwidth}
        \centering
        \includegraphics[width=\textwidth]{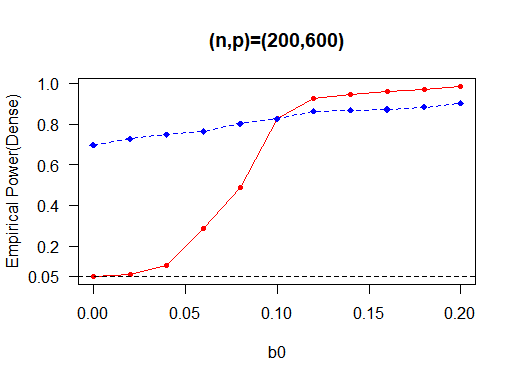}
    \end{subfigure}
    \hspace{0.05\textwidth}
    \begin{subfigure}[b]{0.45\textwidth}
        \centering
        \includegraphics[width=\textwidth]{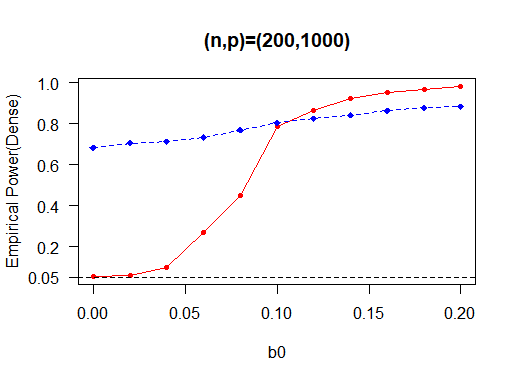}
    \end{subfigure}

    \caption{Empirical sizes and powers for Case 2: the FADS test (solid line) and the test of \cite{jrssb2017} (dashed line).}
    \label{case2}
    \end{figure}

Figures~\ref{case1} and \ref{case2} report the simulation results for two cases as $b_0$ varies. 
Note that when $b_0=0$, the null hypothesis $H_0$ holds, which gives the empirical size.
The case with $b_0>0$ is considered for the local alternative $H_{a1}$.
The departure from $H_{0}$ increases as $b_0$ increases.
For both types of alternative, we observe that the size of the proposed FADS test closely aligns with the nominal level of $0.05$, 
whereas it tends to inflate when employing the method of \cite{jrssb2017}.
Furthermore, our test exhibits higher power for large $b_0$, and approaches 1 as $b_0$ increases in both sparse and dense cases.

In the following analysis, we examine the performance of the proposed test when $\bx_1$ and $\bx_2$ are strongly correlated with each other.
Specifically, we consider the following case for $\bx_1$ and $\bx_2$.

\begin{figure}[thp] 
    \centering
    \begin{subfigure}[b]{0.4\textwidth}
        \centering
        \includegraphics[width=\textwidth]{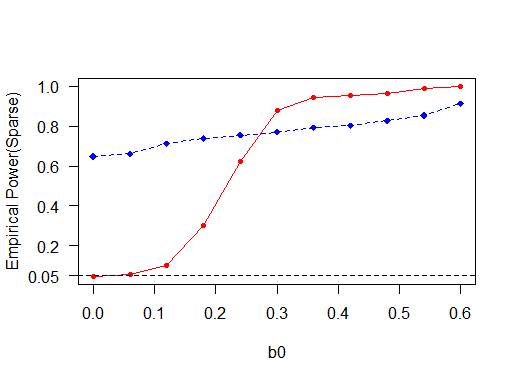}
    \end{subfigure}  
    \hspace{0.05\textwidth}
    \begin{subfigure}[b]{0.4\textwidth}
        \centering
        \includegraphics[width=\textwidth]{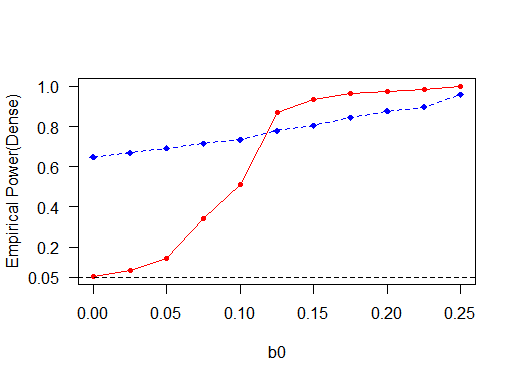}
    \end{subfigure}

    \caption{Empirical sizes and powers for Case 3: the FADS test (solid line) and the test of \cite{jrssb2017} (dashed line).}
    \label{case3}
    \end{figure}

\noindent{\bf Case 3}: We generate $\bx_1$ using $\bx_1=\mathbf P^\top\bx_2+\mathbf\epsilon$, 
where $\bx_2$ and $\mathbf\epsilon$ are independently generated from the mean-zero multivariate normal distribution with the covariance matrix of $(0.5^{|i-j|})_{1\le i,j\le p_2}$ and $(0.5^{|i-j|})_{1\le i,j\le p_1}$.
Here, $\mathbf{P}\in\mathbb{R}^{p_2\times p_1}$ is defined as follows: 
        $$
        \mathbf P=\begin{pmatrix}
            \mathbf 0 &\mathbf 0 &\mathbf 0 \\
            \mathbf Q^\top &\mathbf 0 &\mathbf Q^\top
        \end{pmatrix},
        $$
        where the entries of $\mathbf Q\in \mathbb{R}^{d_1\times d_2}$ are all equal to 0.5. 
       We set $(d_1,d_2)=(5,10)$ and $(n, p)=(200,600)$.
       All the other settings remain as previously described.
Figure~\ref{case3} presents the empirical sizes and powers as \( b_0 \) varies. 
The observations are very similar to those in Cases 1 and 2. 
Specifically, the size of the proposed test is close to the nominal level of $0.05$, 
whereas it tends to inflate for the method of \cite{jrssb2017}.
Furthermore, the power of our test is higher for large $b_0$ and converges to 1 as $b_0$ grows, for both sparse and dense settings.
This indicates that the proposed method remain applicable even when the covariates across different groups are highly correlated.

        \section{Real data analysis} \label{data}

The Cancer Genome Atlas (TCGA) is a landmark project initiated 
by the National Cancer Institute (NCI) and the National Human Genome Research Institute (NHGRI) in the United States. 
It aims to comprehensively characterize genomic alterations across diverse cancer types through large-scale sequencing and analysis. 
TCGA has produced a vast collection of multi-omics data, 
making it a key resource for cancer research. 
The dataset is available at http://cancergenome.nih.gov.

In the TCGA  dataset, gene type refers to the classification of genes based on their functional roles or biological characteristics. 
It describes the type of molecule that a gene encodes (e.g., protein or RNA) and its role in cellular processes. 
Gene type provides important insights into the biological roles of genes in cancer and other diseases, 
and can serve as a natural basis for grouping in our method. 
Here, we study a breast cancer (BRCA) dataset and consider Transcriptome Profiling omics-data. 
Transcriptome Profiling refers to the comprehensive analysis of the transcriptome, 
which is the complete set of RNA transcripts (including mRNA, non-coding RNA, and other RNA molecules) 
produced by the genome in a specific cell or tissue at a given time. 
It provides insights into gene expression patterns, regulatory mechanisms, and functional elements of the genome. 
The sample size is $n=1111$.
There are two gene types in our analysis: protein-coding genes and lncRNA genes. 
Protein-coding genes (examples: TSPAN6, TNMD, DPM1, SCYL3) encode proteins and are the most well-studied functional genes. 
The lncRNA genes (examples: LINC01587, AC000061.1, AC016026.1, IGF2-AS) do not encode proteins 
but are involved in gene regulation and chromatin modification. 
For each subject, there are $p_1=19962$ protein-coding genes and $p_2=16901$ lncRNA genes. 
Figure~\ref{fig3} presents the histogram of sample correlation coefficients computed within and between groups. 
The blue dashed line marks the 75th percentile, highlighting the upper quartile of the correlation distribution. 
Obviously, there is a positive correlation among covariates.

To analyze the protein-coding group, we begin by estimating the number of latent factors using the ratio method as described in  Section~\ref{factor_estimation}, 
which identifies \(\hat{K}_1 = 2\). 
We then apply the FADS test, with the results reported in Table~\ref{table2}. The p-value is 0.018. 
This indicates that the protein-coding group exhibit a significant effect on the survival time conditional on the lncRNA group.
For the lncRNA group, the ratio method identifies \(\hat{K}_2 = 3\). Applying the FADS test gives a p-value of 0.023, see Table~\ref{table2}. This suggests  that the lncRNA group also has a significant effect on the hazard of the survival time after accounting for the protein-coding group.

	\begin{table}[thp]
		\caption{The p-values for each group}
        \label{table2}
		\centering
		
		\begin{tabular}{cccccc}
            \hline
			&Protein-Coding &lncRNA\\
            \hline
            p-value &0.018 &0.023 \\

	 		\hline
	 	\end{tabular}
		
	\end{table}

\section{Discussion} \label{discussion}
 Inspired by the challenges encountered in handling right-censored multi-omics data, this paper presents a FADS test for
  high dimensional group covariates. We established the asymptotic properties of the proposed FADS test under both the null and local hypothesis with estimated latent variables. Numerical simulations have confirmed the effectiveness of our proposed method. Importantly, our test does not require a sparsity assumption on the covariates of the group of interest, thereby broadening the applicability of the proposed method to a wider range of practical problems.

For future research, we aim to investigate the impact of multi-omics data in the framework of functional or partial linear Cox models. Functional data arise in biomedical applications like imaging and longitudinal studies, where covariates vary continuously. Incorporating functional predictors into the Cox model enhances survival analysis by capturing such continuous variations. Meanwhile, when covariate effects on survival outcomes are not purely linear, a partial linear Cox model offers greater flexibility by combining linear and nonparametric components. Developing statistical inference methods for these high-dimensional models presents challenges, particularly in statistical inference and computational efficiency.   

\section*{Supplementary Materials}
The Supplementary Materials includes the proofs of the main theorems along with related technical lemmas.

\section*{Acknowledgments}
Hao's research was partially supported by the Fundamental Research Funds for the Central Universities in UIBE(CXTD14-05), and the National Natural Science Foundation of China(NSFC)(Grant Nos. 12371264 and 12171329).
Qu's research is partially supported by the National Natural Science Foundation of China (12471256) and the Fundamental Research Funds for the Central Universities (No. CCNU25JCPT029).
Sun's research is partially supported by the National Natural Science
Foundation of China (12571299).

\bibliography{ref}  

\end{document}